\documentclass[a4paper,12pt]{article}

\newtheorem{definition}{{\bf Definition}}[section]
\newtheorem{theorem}[definition]{{\bf Theorem}}
\newtheorem{lemma}[definition]{{\bf Lemma}}
\newtheorem{proposition}[definition]{{\bf Proposition}}

\newtheorem{example}[definition]{{\bf Example}}

  \makeatletter
    
    \@addtoreset{equation}{section}
  \makeatother

\def\proof{\noindent{Proof. }}
\def\qed{{\hfill $\square$}\medskip}

\usepackage{amscd,amsmath,amssymb,graphicx,color,mathtools,usebib}

\usepackage{color}

\usepackage[normalem]{ulem}

\begin{document}

\title{A Mathematical Framework for Maze Solving Using Quantum Walks}

\author{Leo Matsuoka\footnote{
Faculty of Engineering, Hiroshima Institute of Technology, Hiroshima 731-5193, Japan;
E-mail: r.matsuoka.65@cc.it-hiroshima.ac.jp}, \quad
Hiromichi Ohno\footnote{
Department of Mathematics, Faculty of Engineering, 
Shinshu University,
4-17-1 Wakasato, Nagano 380-8553, Japan;
E-mail: h\_ohno@shinshu-u.ac.jp}, \quad 
Etsuo Segawa\footnote{
Graduate School of Environment Information Sciences, Yokohama National University,
Yokohama 240-8501, Japan;
E-mail: segawa-etsuo-tb@ynu.ac.jp}
}

\date{}

\maketitle

\noindent {\bf Abstract}

We provide a mathematical framework for identifying the shortest path in a maze using a Grover walk, which becomes non-unitary by introducing absorbing holes. In this study, we define the maze as a network with vertices connected by unweighted edges. Our analysis of the stationary state of the Grover walk on finite graphs, where we strategically place absorbing holes and self-loops on specific vertices, demonstrates that this approach can effectively solve mazes. By setting arbitrary start and goal vertices in the underlying graph, we obtain the following long-time results: (i) in tree structures, the probability amplitude is concentrated exclusively along the shortest path between start and goal; (ii) in ladder-like structures with additional paths, the probability amplitude is maximized near the shortest path.

\section{Introduction}
The development of algorithms for finding the shortest path in networks has a rich history~\cite{network_rev1,network_rev2} and plays an important role in a wide array of practical applications in our daily lives, such as in car navigation systems. The term 'networks' includes various forms, from physical structures like roads and power lines to more abstract ideas. From the study of networks, a larger field known as complex networks has emerged~\cite{complex1,complex2}, which has recently expanded its focus to include quantum mechanics~\cite{complex_q}. The maze-solving problem can be viewed as a fundamental case of the shortest path problem in networks. Here, we define a maze as a network consisting of unweighted edges that represent distance parameters. Consequently, maze-solving techniques are regarded as a subset of the shortest path problem, with various algorithms, including breadth-first search~\cite{book1}, Dijkstra's algorithm~\cite{dijkstra}, and the Bellman–Ford algorithm~\cite{Bellman}, having been developed in classical settings.

The challenge of solving mazes has become an important area of research in the field of natural computing. Various natural phenomena, such as slime molds~\cite{NYT,TKN}, the Belousov–Zhabotinsky (BZ) reaction~\cite{BZ}, electrical discharges~\cite{discharge}, and the propagation of light in waveguides~\cite{photonic}, have mechanisms that can effectively navigate mazes. While humans typically rely on intelligence to solve such problems, nature's ability to accomplish the same task highlights that intelligence is not strictly necessary; instead, the key lies in algorithms governed by simple rules.

In exploring networks like mazes, the movement of a probabilistic walker throughout the network represents a specific type of algorithm. In classical exploration methods, the walker corresponds to probability density, while in quantum exploration methods, it relates to probability amplitude. The basic approach to classical exploration is the random walk. Quantum walks have been introduced as the quantum version of random walks and are expected to improve effectiveness in exploration problems because their probability propagation speed is directly proportional to time, unlike random walks, where it is proportional to the square root of time~\cite{FH,Am1}. Furthermore, quantum walks can utilize the interference of probability amplitudes, providing another advantage in exploration tasks.

The maze-solving approach utilizing quantum walks is derived from quantum search~\cite{Search_K,Search_C,search_A,Portugal}. Investigating specific points in a network using quantum walks is similar to implementing Grover's algorithm on structured networks~\cite{Grover}. Koch et al. extended their research on identifying structural anomalies in networks~\cite{anoma1,anoma2,anoma3} and demonstrated that executing a Grover walk in a maze—where the vertices at the start and goal undergo phase inversion—can result in the temporary identification of the shortest path~\cite{Koch2017,Koch2018,Koch2019,Hillery2021}. In contrast, Matsuoka et al. advanced the study of quantum walks by accounting for both inflow and outflow~\cite{FelHil1,FelHil2,HS,HSS1,HS2,KSS,SKKS}, creating a more robust and versatile approach~\cite{MYLS}. 
In our paper, by integrating self-loop structures at the maze's start and goal, along with incorporating holes that absorb probability amplitudes, we introduce the {\it navigation vector} which is induced by the fundamental cycles and the shortest path between the pair of the vertices having the self-loop of the underlying graph. 
Our mathematical results tells us that a quantum walker takes a time to find autonomously the navigation vector, and as a result, it clings along the shortest path between the start and goal of some mazes in the stationary state. 
 Although Matsuoka et al.'s method is not suitable for mazes with cycles that have an odd number of nodes, it has been shown to extend to tree structures and ladder configurations featuring multiple paths. 
Furthermore, since all time evolution can be expressed as the problem of matrix exponentiation defined by the network structure, this approach enables efficient computations using GPUs, independent of the number of steps or vertices, suggesting its potential as a classical algorithm inspired by quantum mechanics.

In this paper, we present a mathematical reconstruction of the maze-solving method utilizing Grover walk, leading to an explicit derivation of its limit distribution. In Section 2, we outline the detailed structure of the graph that represents the maze and discuss the time evolution of the Grover walk. We also examine the spectral properties of the Grover walk in this context, demonstrating that the probability distribution in the long-time limit is determined by the overlap between the initial state and the eigenspace associated with the eigenvalue $-1$, known as the navigation vector. In Section 3, we detail the limit distribution of the Grover walk specifically for tree structures and ladder graphs.

\section{Settings and properties of Grover walk}

\subsection{Settings}
\bigskip
Let $G_o = (V_o,A_o)$ be a finite connected and symmetric digraph such that an arc $a \in A_o$ if and only if
its inverse arc $\bar{a}\in A_o$.
The origin and terminal vertices of $a \in A_o$ are 
denoted by $o(a) \in V_o$ and $t(a) \in V_o$, respectively.
We assume that $G_o$ has no multiple arcs and self-loops.
This symmetric digraph $G_o$ is called the {\it underlying graph}.

\begin{figure}[h]
\begin{center}
\includegraphics[bb=0 0 600 270, scale=0.4]{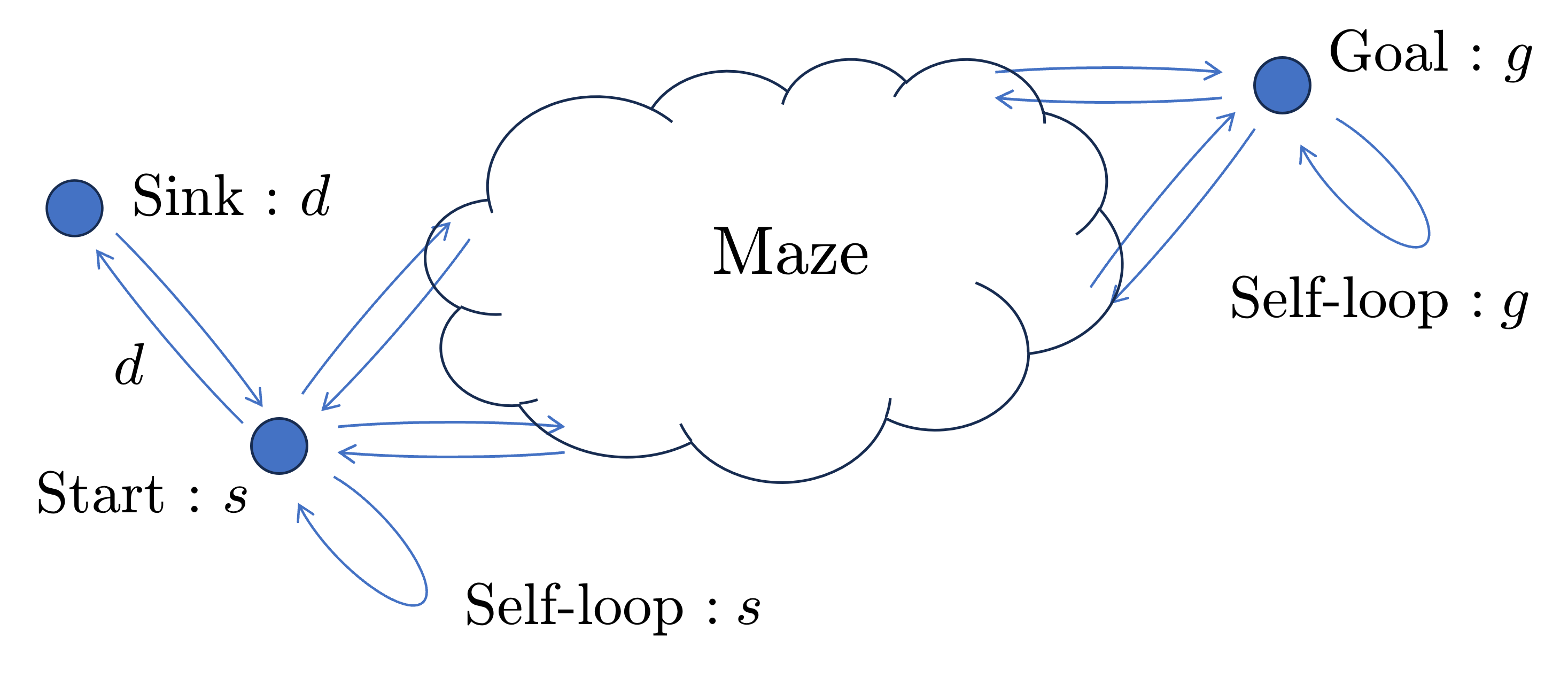}
\vspace*{-0.3cm}
\caption{Setup of the start and goal with self-loops and a sink.}\label{fig1}
\end{center}
\end{figure}

In this paper, we consider the problem of solving a maze by using a quantum walk.
The maze is regarded as the underlying graph $G_o$. To let the quantum walk solve the maze, we slightly deform the underlying graph as follows. 
The start and goal of the maze can be placed at any vertex in $V_o$, and they are indexed by $s$ and $g$, respectively.
To find a route from $s$ to $g$,
we add one self-loop each to vertices $s$ and $g$, which are also indexed by $s$ and $g$.
We also add a vertex which is called sink and indexed by $d$.
The sink is connected only to the start vertex $s$.
The arc which connects $s$ and $d$ is also indexed by $d$, where $o(d) =s$ and $t(d) =d$.
In the following, the graph induced by the underlying graph $G_o$ is denoted by $G=(V,A)$; the digraph $G$ contains these self-loops, the sink vertex and arcs connecting to the start and the sink vertices.
(See Figure \ref{fig1}).
The degree of $v \in V$ is given by
\[
{\rm deg}(v) = |\{ a\in A \, | \, t(a)=v \}|.
\]
By the symmetry of the digraph, the equation ${\rm deg}(v)  =  |\{ a\in A \, | \, o(a)=v \}|$ holds.

The Hilbert space w.r.t. arcs is defined by
\[
{\mathcal H}_A = \left\{ \xi \colon A \to {\mathbb C} \right\}.
\]
We use the Grover walk $U$ on ${\mathcal H}_A$ defined by
\[
(U\xi)(a) = -\xi(\bar{a}) + \frac{2}{{\rm deg}(o(a))} \sum_{t(b)=o(a)} \xi (b)
\]
for any $\xi \in {\mathcal H}_A$.
The initial state is
\[
\zeta (a) = \begin{cases} 1 & (a =s) \\ 0 &(a\neq s) \end{cases}.
\]
We want to assume that the quantum walker at the sink goes away and never comes back.
To represent this, we use the projection $P$ given by
\begin{equation}\label{projP}
(P \xi)(a) = \begin{cases} \xi(a) & a \neq d,\bar{d}\\ 0 & a= d, \bar{d}\end{cases}. 
\end{equation}
For $n=0,1,2,\ldots$, the $n$-th iteration of the Grover walk with the sink is defined by
\[
\zeta_n (a) = \left( (PU)^n \zeta \right) (a) \quad (a \in A).
\]
Then, the finding probability at vertex $v \in V\backslash \{d\}$ and at time $n$ can be defined by
\[
\mu_n(v) = \sum_{t(a)=v} |\zeta_n(a)|^2.
\]

Remark that our assumption is also represented by using an infinite dimensional environmental Hilbert space
${\mathcal H}_E$ and an appropriately extended unitary $\tilde{U}$ on ${\mathcal H}_A \oplus {\mathcal H}_E$ \cite{KSS}.
However, since
\[
(\tilde{U}^n \tilde{\zeta})(a) =  \left( (PU)^n \zeta \right) (a)
\]
for any $a \in A\backslash \{d, \bar{d}\}$,
we will adopt the RHS represented by the finite system in this paper. Here $\tilde{\zeta}$ is the extension of the initial state $\zeta\in \mathcal{H}_A$ to $\mathcal{H}_{A}\oplus \mathcal{H}_E$.

\subsection{Spectral decomposition of $U$ for maze solving}

To solve the maze, it is important to know the eigenvalues and eigenvectors of the Grover walk $U$. 
In particular, we focus on the eigenvectors of $U$ which are invariant under the action of the projection $P$; as we will see, the contribution of the eigenvectors which are not invariant under the action of $P$ reduces exponentially to $0$ as the time step goes to infinity.

We give the following simple lemma which will be useful for the considerations of its remaining statements.
Remark that the eigenvalue $\lambda$ of $U$ satisfies $|\lambda|=1$.

\begin{lemma}\label{lem1.1}
Let $\xi \in {\mathcal H}_A$ be an eigenvector of $U$ associated with the eigenvalue $\lambda$.
Then, $\xi(d) =  \lambda \xi(\bar{d})$.
\end{lemma}
\proof
By the definition of Grover walk, we have
\[
(U \xi)(\bar{d}) = \xi(d).
\]
Then, $U\xi = \lambda \xi$ implies $\lambda \xi(\bar{d}) = \xi(d)$.
\qed

Since our interest is the time iteration of the {\it truncated} Grover walk with respect to $A\setminus\{d,\bar{d}\}$, that is, $PU$, and the support of the initial state is the self-loop $s\in A$ of the start vertex, we will decompose the Grover walk $U$ through the consideration of the overlap between an eigenvector and $\{d,\bar{d},s\}$. The final form of the spectral decomposition can be seen in (\ref{eq:decomposition}). Then, we consider the eigenvectors of $U$.
For an eigenvector $\xi$ of $U$, there are the following four cases:

\smallskip
(1) $\xi(s) \neq 0, \xi(d)=0,$ \qquad  (2) $\xi(s) \neq 0, \xi(d) \neq 0$,

(3) $\xi(s)=0,  \xi (d) \neq 0$, \qquad 
(4) $\xi(s)=0,  \xi(d) = 0$.\\[2mm]
\noindent
Note that by Lemma \ref{lem1.1}, $\xi(d)=0$ implies $\xi(\bar{d})=0$, and $\xi(d)\neq 0$ implies $\xi(\bar{d})\neq 0$.
When a vector $\xi$  satisfies the case $(i)$ $(i\in\{1,2,3,4\})$, we say ``$\xi$ is in $(i)$", for simplicity. 

The eigenspace associated with the eigenvalue $\lambda$ is denoted by $V(\lambda)$.
Note that the eigenvectors in (1) or (4) have the contribution to give the iteration of the truncated Grover walk $PU$, because $d, \bar{d} \notin \mathrm{supp}(\xi)$ for any $\xi$ in (1) or (4). First, we consider the eigenspace $V(-1)$ because of the following strong statement.

\begin{lemma}\label{lem1.2}
Assume that the eigenvector $\xi$ is in {\rm (1)}.
Then, $\xi$ must be an eigenvector of the eigenvalue $-1$.
\end{lemma}
\proof
Let $\{a_i\}_{i=1}^n$ be the set of all arcs 
which satisfy $t(a_i) = s$, where $a_1=\bar{d}$ and $a_2 = s$.
Let $\xi$ have the eigenvalue $\lambda$.
Since $\xi(d)=\xi(\bar{d})=0$ and
\[
\lambda \xi(d) = (U\xi)(d)=-\xi(\bar{d}) + \frac{2}{n} \sum_{i=1}^n \xi(a_i),
\]
we have $\sum_{i=1}^n \xi(a_i) =0$. Therefore,
\[
\lambda \xi(s) = (U\xi)(s) =-\xi(s) + \frac{2}{n} \sum_{i=1}^n \xi(a_i) = -\xi(s).
\]
By the assumption $\xi(s) \neq 0$, we have $\lambda=-1$.
\qed

\begin{figure}[h]
\begin{center}
\includegraphics[bb=0 0 730 180, scale=0.4]{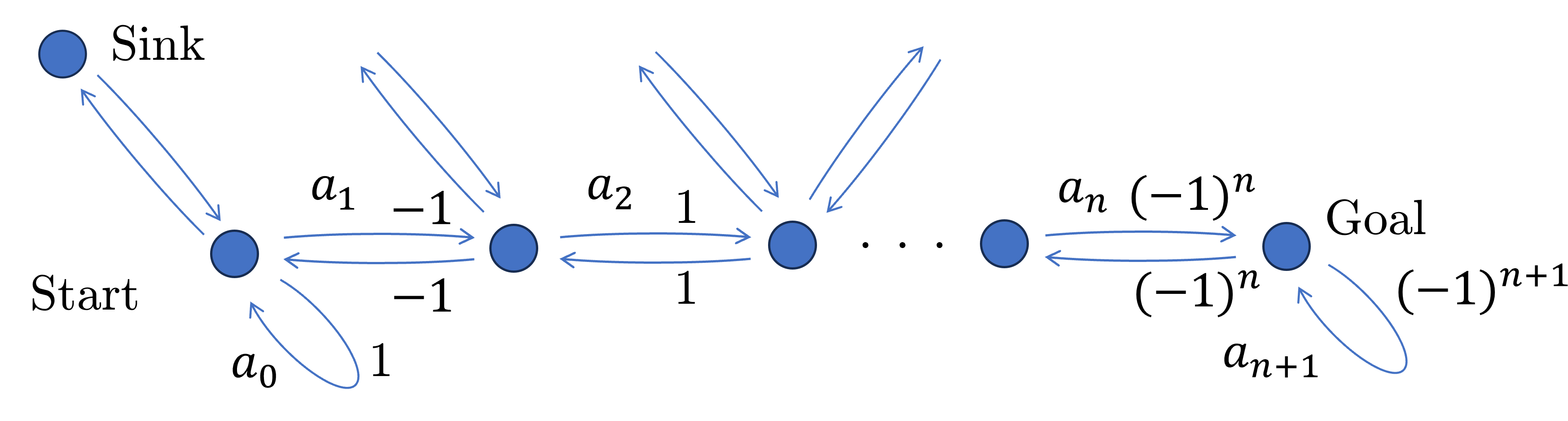}
\vspace*{-0.3cm}
\caption{The eigenvector $\xi\in V(-1)$ with $s\in \mathrm{supp}(\xi)$.} \label{fig2}
\end{center}
\end{figure}

The statement of Lemma~\ref{lem1.2} is strong but not ensures the existence of such an eigenvector. 
However, the following lemma not only ensures such an eigenvector in Lemma~\ref{lem1.2} but also gives an explicit expression by using a path between the self-loops $s$ and $g$.  Figure~\ref{fig2} illustrates this eigenvector.  

\begin{lemma}\label{lem1.3}
There exists an eigenvector  $\xi \in V(-1)$
such that $\xi$ is in {\rm (1)}.
\end{lemma}
\proof 
Let $\{a_i\}_{i=1}^n$ be a path which satisfies $t(a_i)=o(a_{i+1})$, 
$o(a_1)=s$, $t(a_n)=g$, $o(a_i)\neq o(a_j)$ $(i\neq j)$, and $t(a_i)\neq t(a_j)$ $(i\neq j)$,
and let $a_0 =s$ and $a_{n+1}= g$.
Define a vector $\xi \in {\mathcal H}_A$ by
\[
\xi(a) = \begin{cases} (-1)^i  & (a =a_i \ \text{or}\ \bar{a}=a_i) \\ 0 & \text{otherwise} \end{cases}.
\]

\vspace*{-0.5cm}
\noindent
Then, it is easy to check that $\xi$ is an eigenvector in $V(-1)$. (See Figure \ref{fig2}.)\qed

Lemmas~\ref{lem1.2} and \ref{lem1.3} imply that every eigenvector in (1) is included in $V(-1)$. Then, we are interested in whether $V(-1)$ includes eigenvectors in (2), (3) or (4). Let us see that the following lemma removes the possibilities that $\xi\in V(-1)$ is in  (2) or (3). 

\begin{lemma}\label{lem1.4}
Every eigenvector $\xi$ in $V(-1)$ satisfies $\xi(d)=0$, that is , $\xi$ is in {\rm (1) or (4)}.
\end{lemma}
\proof
Let $\xi$ be an eigenvector in $V(-1)$.
We use the same setting as in the proof of Lemma \ref{lem1.2}.
Similar to the proof of Lemma \ref{lem1.2}, 
\[
-\xi(s) = (U\xi)(s) = -\xi(s) +\frac{2}{n}\sum_{i=1}^n \xi(a_i).
\]
Hence, $\sum_{i=1}^n \xi(a_i) =0$. Since $\xi(d) = -\xi(\bar{d})$ by Lemma \ref{lem1.1},
\[
-\xi(d) = (U\xi)(d) = -\xi(\bar{d}) +\frac{2}{n}\sum_{i=1}^n \xi(a_i) = -\xi(\bar{d}) =\xi(d).
\]
This implies $\xi(d) =0$. 
\qed

Now we are ready to construct an ONB for $V(-1)$ as follows. 

\begin{proposition}\label{prop1.5}
There exists an ONB  $\{ \psi^{(-1)}_1, \ldots, \psi^{(-1)}_{\dim V(-1) -1},\varphi \}$ of $V(-1)$
such that $\varphi$ is in {\rm (1)} and $\psi_i^{(-1)}$ is in ${\rm (4)}$.
\end{proposition}
\proof
Let $k = {\dim V(-1)}$, and let $\{\xi_1 , \ldots, \xi_k\}$ be a bases of $V(-1)$ with $\xi_k(s)\neq 0$. 
We can take such a basis by setting $\xi_k$ to be in (1).
Define a vector $\eta_i$ by
\[
\eta_i = \begin{cases} \displaystyle \xi_i - \frac{\xi_i(s)}{\xi_k(s)} \xi_k & (1\le i \le k-1) \\ 
\xi_k & (i=k) \end{cases}.
\]
Remark that $\eta_i$ is in (4) for $1\le i \le k-1$.
We apply the Gram-Schmidt process to $\{\eta_i\}_{i=1}^k$, and denote the constructed vectors by
$\{\psi_1^{(-1)}, \ldots, \psi_{k-1}^{(-1)}, \varphi\}$. This is the desired ONB. \qed

The reason for choosing such an ONB of $V(-1)$ is as follows. 
To describe the Fourier expansion of the initial state $\zeta$, we use the inner products of each vector in an ONB and $\zeta$. 
Note that the first $\mathrm{dim} V(-1)-1$ bases of $V(-1)$ in  Proposition~\ref{prop1.5}, $\psi_j^{(-1)}$ ($1\le j \le \mathrm{dim}V(-1)-1$), have no overlap to the initial state $\zeta$, that is, $\langle \psi_j^{(-1)}, \zeta\rangle=0$ for any $j=1,\dots,\mathrm{dim}V(-1)-1$. The only base which has the overlap with $\zeta$ is the last base $\varphi$. Thus the overlap of the initial state with the subspace $V(-1)\subset \mathcal{H}_A$ is obtained by just one inner product of $\varphi$ and $\zeta$ by this expression of ONB. 
As we will see, the contribution of $V(\lambda)$ with $\lambda\neq -1$ to the survival probability $\|(PU)^n \zeta\|^2$ exponentially reduces. 
Thus the vector $\varphi$ is the key to {\it navigates} the quantum walker to the stationary state. 

\bigskip

Next, we are interested in the contribution of the eigenspace of $U$ having the overlap with the initial state for the eigenvalue $\lambda\neq -1$ to the survival probability. To this end, we construct an ONB of $V(\lambda)$ with $\lambda \neq -1$ step by step; the final form of the decomposition of $U$ for the survival probability is described in \eqref{eq:decomposition}.
First, we show that every eigenvector with the eigenvalue $\lambda\neq -1$ is not in (2), or is not in (3) as follows. 

\begin{lemma}\label{lem1.6}
Suppose $\lambda \neq -1$. $V(\lambda)$ does not contain both of vectors in {\rm (2)} and {\rm (3)}.
\end{lemma}
\proof
When $\xi_2 \in V(\lambda)$ is in (2) and 
$\xi_3 \in V(\lambda)$ is in (3),
we can construct an eigenvector $\xi_1 \in V(\lambda)$ in (1) using a linear combination.
Then, $\lambda =-1$ by Lemma \ref{lem1.2}, which is contradiction. \qed

Using Lemma~\ref{lem1.6}, we can construct the following ONB for $\lambda\neq -1$ concerning to the overlap with $d$ and $\bar{d}$. 
\begin{proposition}
There exists an ONB $\{\psi^{(\lambda)}_1, \ldots, \psi_{\dim V(\lambda)}^{(\lambda)}\}$ 
of $V(\lambda)$
such that at most one vector in the ONB is in {\rm (2)} or {\rm (3)}.
\end{proposition}
\proof
If there is no vector in (2) or (3) in $V(\lambda)$, then all vectors in $V(\lambda)$ are in (4),
and we can construct the desired ONB, easily.

If there is a vector in (2) or (3), we can use the same method in the proof of Proposition \ref{prop1.5}. Assume that there is a vector in (2).
Let $k = {\rm dim}V(\lambda)$, and let $\{\xi_1,\ldots, \xi_k\}$ be a basis of $V(\lambda)$
with $\xi_k (s) \neq 0$ and $\xi_k(d) \neq 0$.
Define a vector $\eta_i$ by
\[
\eta_i = \begin{cases} \displaystyle \xi_i - \frac{\xi_i(d)}{\xi_k(d)} \xi_k & (1\le i \le k-1) \\ 
\xi_k & (i=k) \end{cases}.
\]
Remark that $\eta_i$ is in (4) for $1\le i \le k-1$, because $\eta_i(d) =0$ and there is no vector in (1).
We apply the Gram-Schmidt process to $\{\eta_i\}_{i=1}^k$, and denote the constructed vectors by
$\{\psi_1^{(\lambda)}, \ldots, \psi_{k}^{(\lambda)} \}$. This is the desired ONB. 

When there is a vector in (3), we can construct the desired ONB, similarly.
\qed

We denote $k^{(\lambda)}$ by the number of vectors in (4) in the above ONB, and put
$k^{(-1)} = {\rm dim}V(-1)-1$.
When the above ONB contains a vector in (2) or (3), we denote the vector by
$\eta^{(\lambda)}$. This means $\psi_{k^{(\lambda)}+1} = \eta^{(\lambda)}$.

\bigskip

Here, we consider the spectrum decomposition of $U$, that is,
\begin{equation}\label{eq:decomposition}
U = -|\varphi \rangle \langle \varphi | +
\sum_{\lambda \in \sigma(U)}  \lambda \left( |\eta^{(\lambda)}\rangle \langle \eta^{(\lambda)}|
+ \sum_{i=1}^{k^{(\lambda)}}  |\psi^{(\lambda)}_i \rangle \langle \psi_i^{(\lambda)}| \right),
\end{equation}
where $\sigma(U)$ is the set of all eigenvalues of $U$, 
and $\eta^{(\lambda)}=0$ when $V(\lambda)$ does not contain a vector in (2) or (3).
Define
${\mathcal K} = {\rm span} \{ \eta^{(\lambda)} \, | \, \lambda\in \sigma(U) \}$.
We see that this subspace is invariant under $P$ defined in \eqref{projP}.

\begin{lemma}\label{lem1.8}
${\mathcal K}$ is invariant under $P$, that is, $P{\mathcal K} \subset {\mathcal K}$.
\end{lemma}
\proof
Remark that $P\varphi =\varphi$ and $P\psi_i^{(\lambda)} = \psi_i^{(\lambda)}$
$(1\le i \le k^{(\lambda)})$.
It is sufficient to prove that $P{\mathcal K}$ is orthogonal to $\varphi$ and $\psi_i^{(\lambda)}$.
For $\xi \in {\mathcal K}$, we have
\[
\langle \varphi, P\xi \rangle = \langle P \varphi, \xi \rangle = \langle \varphi, \xi \rangle =0.
\]
Similarly, we can check that $P\xi$ is orthogonal to $\psi_i^{(\lambda)}$.   \qed

This Lemma implies that the contribution of the subspace $\mathcal{K}\subset \mathcal{H}_A$ to the survival probability reduces exponentially to zero as time goes to infinity as follows. 
\begin{proposition}
A vector $\xi \in {\mathcal K}$ satisfies $\lim_{m\to \infty} (PU)^m \xi = 0$.
\end{proposition}
\proof
Since ${\mathcal K}$ is invariant under $PU$ by Lemma \ref{lem1.8},
it is enough to prove that the absolute value of the eigenvalue of $PU|_{{\mathcal K}}$ is less than $1$.

Assume that $\lambda$ is an eigenvalue of $PU |_{{\mathcal K}}$ with $|\lambda|=1$, and
$\eta$ is a unit eigenvector associated with the eigenvalue $\lambda$.
The equation
\[
\| PU \eta\| =\|\lambda \eta\|= \|\eta\| = \|U\eta\|
\]
implies $ (U\eta)(d) = (U\eta)(\bar{d}) =0$, and therefore, $PU\eta = U\eta$.
Hence, we have
\[
U\eta = PU\eta = \lambda \eta,
\]
which means that $\eta$ is an eigenvector of $U$.
Moreover, $\eta(d) = 0$.

Since $\eta$ has the eigenvalue $\lambda$, $\eta$ is in $V(\lambda)$. 
On the other hand, $\eta$ is in ${\mathcal K}$.
Hence, $\eta$ is equal to $e^{{\rm i}t} \eta^{(\lambda)}$ for some $t \in {\mathbb R}$.
However, this contradicts $\eta(d)=0$.
\qed

The initial state $\zeta$ is written as
\[
\zeta = \langle \varphi,\zeta\rangle \varphi + 
\sum_{\lambda \in \sigma(U)} \langle \eta^{(\lambda)} , \zeta\rangle \eta^{(\lambda)}
=\overline{\varphi(s)} \varphi + \sum_{\lambda \in \sigma(U)} 
\overline{\eta^{(\lambda)}(s)}\eta^{(\lambda)}.
\]
So, 
\[
\lim_{n\to \infty}(-1)^n (PU)^n \zeta 
= \overline{\varphi(s)} \varphi.
\]
We will call this vector $\varphi$ the {\it navigation} vector.
Recall that the navigation vector is the unique base of $V(-1)$ which has an overlap to $s$, see Proposition~\ref{prop1.5}. 
We use this navigation vector to solve the maze because the survival probability can be simply described by only the overlap of the initial state and the navigation vector:
\begin{proposition}\label{prop:survivalprob}
Let $\mu_n(v)$ be the probability at position $v$ at time $n$, and $\varphi$ be the navigation vector defined the above. Then we have 
\[ \lim_{n\to\infty}\mu_n(v)= |\varphi(s)|^2\sum_{t(a)=v}|\varphi(a)|^2. \]
\end{proposition}
Therefore, finding the ONB in Proposition~\ref{prop1.5} so that the navigation vector appears is the key to obtain the survival probability. 
We demonstrate the construction of the ONB and the navigation vector for a tree case and a ladder case in the next section.

\section{Maze solving}

\subsection{Tree case}\label{sectTree}

When the graph is a tree, we can determine the navigation vector.

\begin{theorem}\label{thm:tree}
When the underlying graph is a tree, the navigation vector is constant multiple of the vector defined in the proof of Lemma {\rm \ref{lem1.3}}.
\end{theorem}
\proof
When the underlying graph $G_o=(V_o,E_o)$ is bipartite, the dimension of $V(-1)$ is given by $(|E_o|-|V_o|+1)+1$~\cite[Proposition~6 for Case (C)]{KSS}. Note that for tree case, the Betti number $|E_o|-|V_o|+1$ becomes $0$. Then $\dim V(-1)=1$. 
On the other hand, the navigation vector and the vector defined in Lemma \ref{lem1.3} are both in $V(-1)$. 
These imply the assertion of this theorem. \qed

When the underlying graph is a tree, the shortest route from the start to the goal is uniquely determined
which is denoted by $\{a_i\}_{i=1}^n$. 
We set $a_0 =s$ and $a_{n+1}=g$.
Let $\varphi$ be navigation vector, and let $r = |\varphi(s)|$.
Then, the navigation vector satisfies
\[
|\varphi (a)| = \begin{cases} r & a = a_i \ {\rm or}\ \bar{a} =a_i \\ 0 & {\rm otherwise} \end{cases}.
\]
This means that the amplitudes of the navigation vector are only on the shortest route.
Therefore, the position of the quantum walker tells us the shortest route.

\subsection{The case the graph has cycles}\label{sect4}

Note that a tree has the unique simple path between the start and goal vertices, and this becomes the shortest path. Here, a simple path is a walk in which all vertices and also all edges are distinct. 
We have seen in Section~\ref{sectTree} that the shortest path of arbitrary tree is only the place where a quantum walker ``clings"  in the long time limit. 
On the other hand, once there is a cycle, there are several choices of the non-backtracking path. As a natural question, we are interested in the tendency of a quantum walker in this case. 
In this section, we consider a ladder graph as an example of the graph which has cycles so that there are $k+1$ kinds of ``detours" among the shortest path. See Figure~\ref{fig6}. 
Before describing the setting, we prove the next lemma.

\begin{figure}[h]
\begin{center}
\includegraphics[bb=0 0 520 180, scale=0.4]{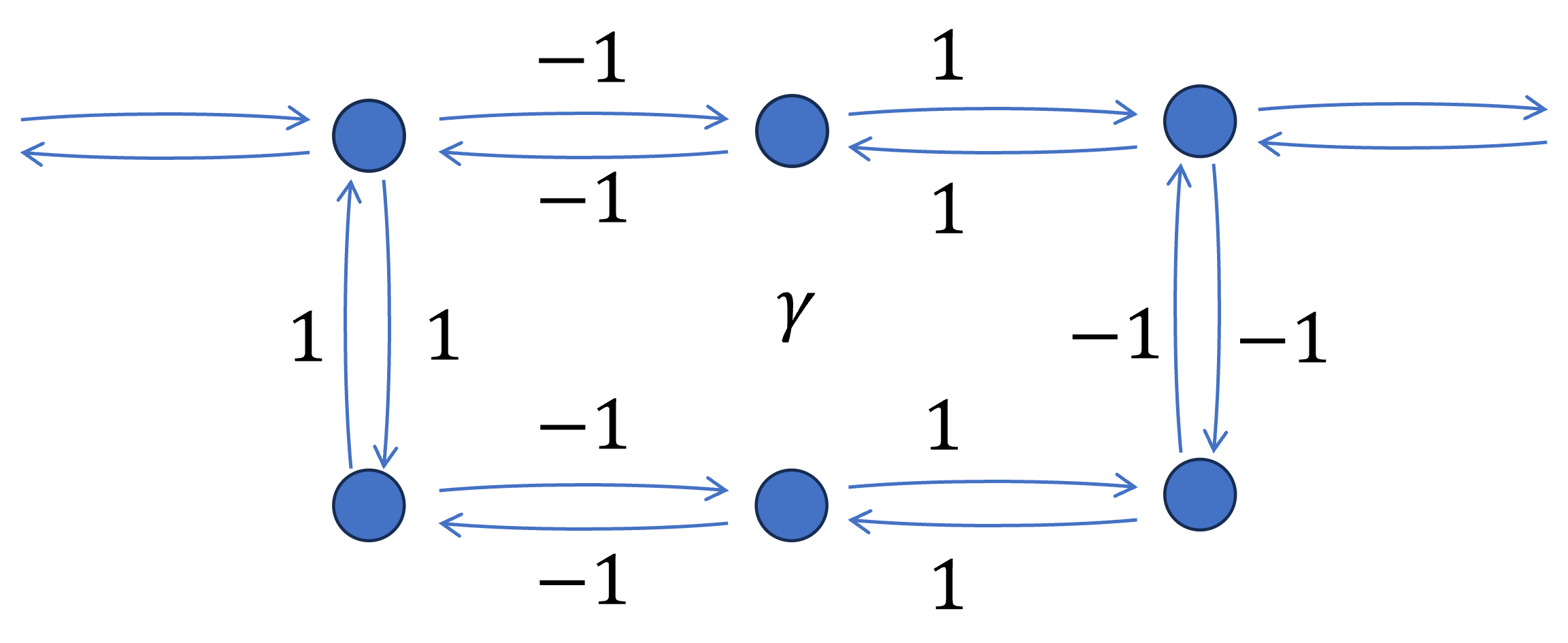}
\vspace*{-0.3cm}
\caption{An example of an eigenvector associated with a cycle.}\label{fig5}
\end{center}
\end{figure}

\begin{figure}[h]
\begin{center}
\includegraphics[bb=0 0 920 350, scale=0.4]{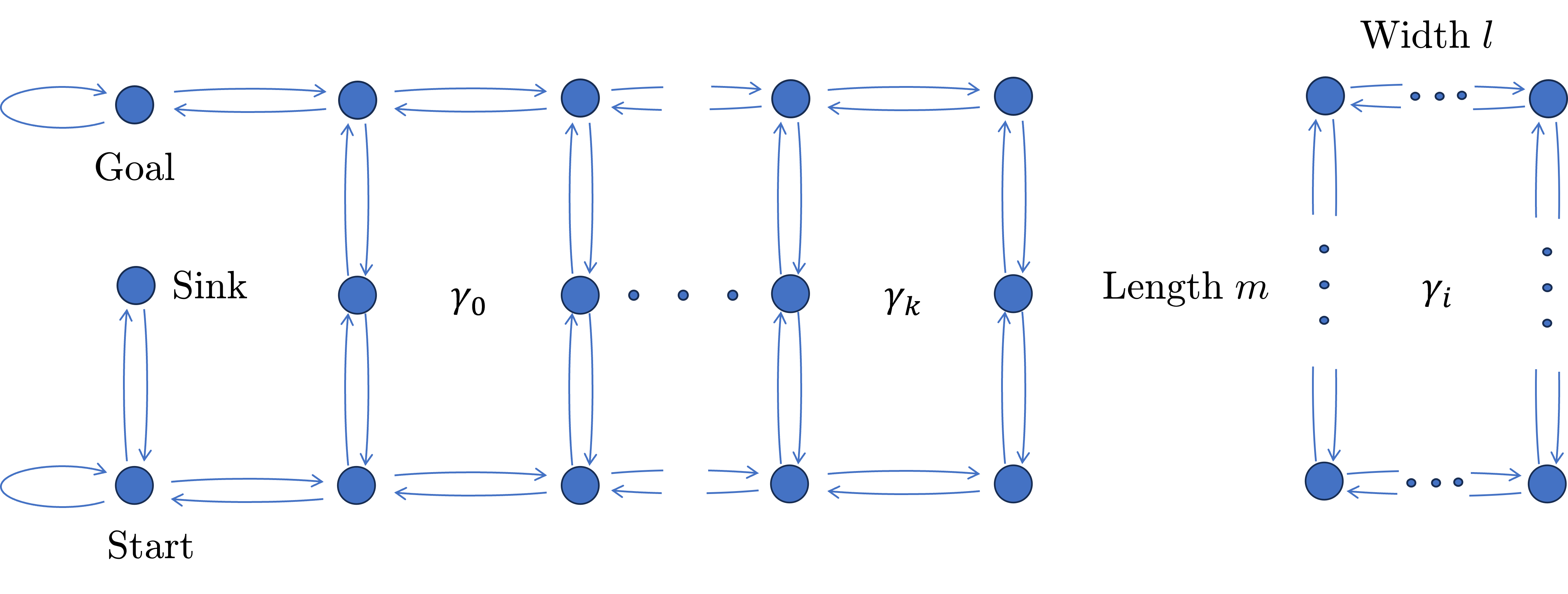}
\vspace*{-0.3cm}
\caption{The ladder graph with the self loops and the sink.  The left figure illustrates the graph treated in Section~\ref{sect4}. This graph is constructed by $k+1$ cycles $\gamma_i$ ($i=0,1,\dots,k$) depicted by the right figure. Each cycle whose length is $m$ and width is $\ell$ has the eigenvector of the eigenvalue $-1$, $\gamma_i$ ($i=0,\dots,k$). (The notations for the cycle and the induced eigenvector are same. )
}\label{fig6}
\end{center}
\end{figure}

\begin{lemma}\label{lem4.1}
For a cycle $\gamma \in G$ of length $2n$ $(n \in {\mathbb N})$, 
there exists an eigenvector of the Grover walk $U$ associated with the eigenvalue $-1$.
\end{lemma}
\proof
Let $\{a_i\}_{i=1}^{2n}$ be a cycle of length $2n$, that is, $t(a_i)=o(a_{i+1})$, $o(a_1) = t(a_{2n})$ and $t(a_i) \neq t(a_j)$ $(i\neq j)$.
Define a vector $\xi \in {\mathcal H}_A$ by
\[
\xi(a) = \begin{cases} (-1)^i & (a=a_i \ {\rm or} \ \bar{a} =a_i) \\ 0 & {\rm otherwise} \end{cases}.
\]
This is the desired eigenvector. (See Figure \ref{fig5}.) \qed

Now, we consider a ladder graph. 
As in Figure \ref{fig6}, the ladder graph contains $k+1$ cycles which are indexed by $\gamma_i$.
Each cycle $\gamma_i$ is a rectangle of length $m$ and width $l$.
The lower left vertex of the cycle $\gamma_0$ and the start vertex are adjacent.
The normalized  eigenvector with respect to the cycle $\gamma_n$ 
constructed in Lemma \ref{lem4.1} is also
denoted by $\gamma_n$ $(0 \le n \le k)$, and
the normalized eigenvector with respect to the shortest route defined in  Lemma \ref{lem1.3} 
is denoted by $\xi$.
Without loss of generality, we can assume $\langle \gamma_n, \gamma_{n+1}\rangle \ge 0$ and
$\langle \gamma_0, \xi \rangle \ge 0$.
It is known that the eigenvector of $U$ associated with the eigenvalue $-1$ is described 
as a linear combination of $\gamma_n$ and $\xi$ \cite{KSS}.
Indeed, it holds that $\mathrm{span}\{\gamma_0\dots,\gamma_{k}\}=\mathrm{span}\{\psi_1^{(-1)},\dots,\psi_{k^{(-1)}}^{(-1)} \}$, where $\psi_i^{(-1)}$'s are the eigenvectors of the eigenvalue $-1$ in (\ref{eq:decomposition}) whose supports have no
intersection to $\{d,\bar{d}\}$. To obtain the navigation vector $\varphi$ in (\ref{eq:decomposition}), from now on, we construct the ONB of $V(-1)$  in Proposition~\ref{prop1.5} by applying the Gram-Schmidt process to $\{\gamma_0,\dots,\gamma_k,\xi\}$ step by step. 

\begin{lemma}\label{lemma4.2}
The equations
\begin{align*}
\langle \gamma_s, \xi\rangle =0 \quad (s \ge 1), \qquad
\langle \gamma_s, \gamma_t \rangle =0 \quad (|s-t|\ge 2), \qquad 
\langle \gamma_s, \gamma_{s+1} \rangle = \frac{m}{2(l+m)}
\end{align*}
hold.
\end{lemma}
\proof
The first and second equations are obvious by the definition of $\gamma_s$ and $\xi$.
The inner product in the third equation can be calculated as
\[
\langle \gamma_s , \gamma_{s+1} \rangle = \frac{1}{4(l+m)} \sum_{i=1}^m 2 =\frac{m}{2(l+m)},
\]
because of the assumption $\langle \gamma_s, \gamma_{s+1}\rangle \ge 0$ \qed

Let $\kappa = \frac{m}{2(l+m)}$. 
We apply the Gram-Schmidt process to $\{ \gamma_0, \ldots, \gamma_k, \xi \}$.
Since $\gamma_n$ is in (4) and $\xi$ is in (1), 
the constructed ONB is the ONB in Proposition \ref{prop1.5}.
We denote this by $\{\psi_0, \ldots, \psi_k, \varphi \}$.
Then, $\varphi$ is the navigate vector.

First, we consider about $\gamma_n$ and $\psi_n$.
Let $U_n(x)$ be the second kind of Chebyshev polynomial, 
and define $u_n = U_n\left(\frac{1}{2\kappa}\right)$.

\begin{proposition}\label{prop4.3}
The equation
\[
\psi_n =  \frac{1}{\sqrt{\kappa u_n u_{n+1}}} \sum_{s=0}^n (-1)^{n-s} u_s\gamma_s
\]
holds.
\end{proposition}

\proof
By $\psi_s \in {\rm span}\{\gamma_0, \ldots, \gamma_s\}$ and Lemma \ref{lemma4.2},
\[
\langle \psi_s , \gamma_t \rangle =0
\]
for $t \ge s+2$. 
Since ${\rm span}\{\gamma_0, \ldots, \gamma_n\} = {\rm span}\{\psi_0, \ldots, \psi_n\}$,
$\gamma_n = \sum_{i=0}^n \langle \psi_i, \gamma_n \rangle \psi_i$.
These equations imply $\gamma_n \in {\rm span}\{\psi_{n-1}, \psi_n\}$, where $\psi_{-1}=0$. 
Here, we put
\[
\gamma_n = a_n \psi_{n-1} + b_n \psi_n,
\]
where $a_0=0$. Remark that $b_0=1$ and $a_n, b_n \in {\mathbb R}$.
Since $\psi_n$ is the normalization of 
$\gamma_n - \sum_{s=0}^{n-1} \langle \psi_s, \gamma_n\rangle \psi_s$ and
$\langle \psi_n, \gamma_n\rangle = b_n $, $b_n$ is positive.
Moreover, 
\[
\langle \gamma_n, \gamma_{n+1} \rangle 
= \langle a_n \psi_{n-1} + b_n \psi_n,a_{n+1} \psi_{n} + b_{n+1} \psi_{n+1} \rangle
= a_{n+1}b_n
\]
implies $a_{n+1}b_n = \kappa$, and $\|\gamma_n\|=1$ implies $a_n^2+b_n^2=1$.
The equation $a_{n+1}b_n =\kappa$ means that $a_{n+1}$ and $b_n$ have the same sign,
and therefore, $a_n$ is positive for $n \ge 1$.
To calculate $a_n^2$ and $b_n^2$, we use Chebyshev polynomial.
We claim that
\[
a_n^2 = \frac{\kappa u_{n-1}}{u_n}, \qquad b_n^2 = \frac{\kappa u_{n+1}}{u_n},
\]
where $u_{-1}=0$.
Indeed, we can prove this using a similar method to the proof in \cite{SKKS}, 
or we can check that these $a_n^2$ and $b_n^2$ satisfy the above recurrence relations.

By this fact and $\gamma_n = a_n\psi_{n-1} + b_n\psi_{n}$, we have
\begin{align*}
\psi_n &= \frac{1}{b_n} ( \gamma_n - a_n\psi_{n-1})
= \frac{1}{b_n} \left( \gamma_n -\frac{a_n}{b_{n-1}} (\gamma_{n-1} - a_{n-1}\psi_{n-2}) \right)
= \cdots \\
&=\frac{1}{b_n} \sum_{s=0}^n (-1)^{n-s} \prod_{j=s+1}^n \frac{a_j}{b_{j-1}} \gamma_s
= \frac{1}{\sqrt{\kappa u_n u_{n+1}}} \sum_{s=0}^n (-1)^{n-s} u_s\gamma_s
\end{align*}
which is the assertion. \qed

Next, we consider about $\xi$ and $\varphi$. 
Define a projection $\Pi_k$ by
\[
\Pi_k = \sum_{n=0}^k |\psi_n\rangle \langle \psi_n |.
\]
$\Pi_k$ is also the projection onto ${\rm span}\{\gamma_0, \ldots, \gamma_k\}$.
Here, $\varphi$ is written as
\[
\varphi = \frac{(I-\Pi_k) \xi}{ \| (I-\Pi_k)\xi\|}.
\]

\begin{lemma}\label{lemma4.4}
\[
\|(I-\Pi_k) \xi \|^2 = 1 - \frac{m}{m+3} \sum_{n=0}^k \frac{1}{u_n u_{n+1}}.
\]
\end{lemma}

\proof
By the Pythagorean theorem, 
\[
\|(I-\Pi_k) \xi \|^2 = \|\xi\|^2 - \|\Pi_k \xi\|^2  = 1 - \|\Pi_k \xi\|^2.
\]
Since 
\[
\langle \gamma_0, \xi\rangle = \frac{m}{\sqrt{2(l+m)(m+3)}},
\]
we have 
\begin{align*}
\|\Pi_k \xi \|^2 &=\sum_{n=0}^k |\langle \psi_n , \xi \rangle |^2
=\sum_{n=0}^k \left| \left\langle 
\frac{1}{\sqrt{\kappa u_n u_{n+1}}} (-1)^n u_0 \gamma_0, \xi \right\rangle \right|^2 \\
&=\sum_{n=0}^k \frac{1}{\kappa u_n u_{n+1}} \cdot \frac{m^2}{2(l+m)(m+3)} \\
&=\frac{m}{m+3} \sum_{n=0}^k \frac{1}{u_n u_{n+1}}
\end{align*}
by Lemma \ref{lemma4.2} and Proposition \ref{prop4.3}. \qed

\begin{proposition}\label{prop3.6.1}
The navigation vector $\varphi$ is written as 
\[
\varphi = \left( \xi - \frac{\sqrt{2(l+m)}}{\sqrt{m+3}} \sum_{n=0}^k \frac{1}{u_n u_{n+1}}
\sum_{s=0}^n (-1)^s u_s \gamma_s \right) \big/ \|(I-\Pi_k)\xi\|.
\]
\end{proposition}

\proof
By Lemma \ref{lemma4.2} and Proposition \ref{prop4.3},
\begin{align*}
(I-\Pi_k)\xi &= \xi - \Pi_k \xi 
=
\xi - \sum_{n=0}^k \langle \psi_n ,\xi \rangle \psi_n \\
&=
\xi - \sum_{n=0}^k \frac{1}{\sqrt{\kappa u_n u_{n+1}}}(-1)^n   \frac{m}{\sqrt{2(l+m)(m+3)}} \psi_n\\
& = \xi -  \sum_{n=0}^k \frac{1}{\sqrt{\kappa u_n u_{n+1}}}(-1)^n   \frac{m}{\sqrt{2(l+m)(m+3)}} 
\cdot  \frac{1}{\sqrt{\kappa u_n u_{n+1}}} \sum_{s=0}^n (-1)^{n-s} u_s\gamma_s \\
&= \xi - \frac{\sqrt{2(l+m)}}{\sqrt{m+3}} \sum_{n=0}^k \frac{1}{u_n u_{n+1}} 
\sum_{s=0}^n (-1)^s u_s \gamma_s
\end{align*}
which is the assertion. \qed

By this proposition, we can calculate $\varphi(a)$ for an arc $a$, explicitly.
Let $a$ be an arc on the left side of the cycle $\gamma_0$.
If $\gamma_0(a) >0$, $\gamma_0(a) = \frac{1}{2\sqrt{m+l}}$.
Moreover, by the assumption $\langle \gamma_0, \xi \rangle \ge 0$, $\xi(a) >0$.
Hence, when $\gamma_0(a) >0$, we have
\begin{align*}
\|(I-\Pi_k)\xi\| \cdot \varphi(a) &= 
\xi(a) - \frac{\sqrt{2(l+m)}}{\sqrt{m+3}} \sum_{n=0}^k \frac{1}{u_n u_{n+1}}
\sum_{s=0}^n (-1)^s u_s \gamma_s(a) \\
&=\frac{1}{\sqrt{2(m+3)}} -\frac{\sqrt{2(l+m)}}{\sqrt{m+3}} 
\sum_{n=0}^k \frac{1}{u_n u_{n+1}} \frac{1}{2\sqrt{l+m}} \\
&=
\frac{1}{\sqrt{2(m+3)}} \left( 1 - \sum_{n=0}^k \frac{1}{u_n u_{n+1}} \right).
\end{align*}
When $\gamma_0(a)<0$, just multiply the above equation by $-1$. 
Similarly, when $a$ is an arc on the left side of the cycle $\gamma_i$ for $1\le i \le k$ 
and $\gamma_i(a) >0$,
\begin{align*}
&\|(I-\Pi_k)\xi\| \cdot \varphi(a) = 
\xi(a) - \frac{\sqrt{2(l+m)}}{\sqrt{m+3}} \sum_{n=0}^k \frac{1}{u_n u_{n+1}}
\sum_{s=0}^n (-1)^s u_s \gamma_s(a)\\
&\quad =
- \frac{1}{\sqrt{2(m+3)}} 
\left( \frac{1}{u_{i-1}u_i} (-1)^{i-1} u_{i-1} +
\sum_{n=i}^k \frac{1}{u_n u_{n+1}} ((-1)^{i-1} u_{i-1} + (-1)^i u_i)\right).
\end{align*}
When $\gamma_i(a)<0$, just multiply the above equation by $-1$. 

In order to compare these values, we prepare the next lemma.

\begin{lemma}\label{lem4.6}
The following inequalities hold for $i \ge 0$:\\
$\quad (1)  \quad u_{i+1} > u_i +1,$ \\
$\quad (2)  \quad u_{i+2}-u_{i+1} > u_{i+1} - u_{i},$ \\
$\quad (3)  \quad \displaystyle \sum_{n=i}^k \frac{1}{u_nu_{n+1}} < \frac{1}{u_i} \cdot \frac{1}{u_{i+1} - u_i}.$
\end{lemma}

\proof (1)
Let $x = \frac{1}{2\kappa}$. When $n=0$, $u_0 = U_0(x) = 1$ and $u_1 =U_1(x) = 2x$.
Since $x >1$, we have $u_1 \ge u_0+1$. If the inequality is true for $i$, we have
\[
u_{i+2} = 2x u_{i+1} -u_i = u_{i+1} + (2x-2)u_{i+1} + (u_{i+1}-u_i) > u_{i+1} +1.
\]
Hence, we obtain the assertion by induction.

(2) We can calculate as
\[
u_{i+2} - u_{i+1} = (2x-1) u_{i+1} - u_{i} > u_{i+1}-u_i.
\]

(3) By using the partial fraction decomposition,
\begin{align*}
&\sum_{n=i}^k \frac{1}{u_nu_{n+1}} = 
\sum_{n=i}^k \frac{1}{u_{n+1} -u_n} \cdot \left( \frac{1}{u_n} - \frac{1}{u_{n+1}}\right) \\
&= \frac{1}{u_i} \cdot \frac{1}{u_{i+1}-u_i} -\frac{1}{u_{k+1}} \cdot \frac{1}{u_{k+1}-u_k}
-\sum_{n=i}^{k-1} \frac{1}{u_{n+1}} \left( \frac{1}{u_{n+1} -u_n} - \frac{1}{u_{n+2} -u_{n+1}} \right)\\
&< \frac{1}{u_i} \cdot \frac{1}{u_{i+1}-u_i}.
\end{align*}
We use (2) for the last inequality. \qed

The next proposition shows that 
the probability $p_i$ of finding the quantum walker at the left hand side of $\gamma_i$ 
is monotonically decrease.

\begin{theorem}\label{prop3.8.1}
Let $a_i$ be an arc on the left side of the cycle $\gamma_i$ for $0\le i \le k$.
Then, the inequality
\[
|\varphi(a_i)|>|\varphi(a_{i+1})|
\]
holds for $0 \le i \le k-1$.
\end{theorem}

\proof
First, we show that 
\[
\frac{1}{u_i} +(u_{i-1} -u_i) \sum_{n=i}^k \frac{1}{u_nu_{n+1}}
\]
is positive. Indeed, by Lemma \ref{lem4.6} (2) and (3),
\[
\frac{1}{u_i} +(u_{i-1}-u_i) \sum_{n=i}^k \frac{1}{u_nu_{n+1}} 
> \frac{1}{u_i} - (u_i-u_{i-1}) \cdot \frac{1}{u_i} \cdot \frac{1}{u_{i+1}-u_{i}} >0.
\]

When $i=0$, it is enough to show 
\[
1 -\sum_{n=0}^k \frac{1}{u_nu_{n+1}} > \frac{1}{u_1} +(1-u_1) \sum_{n=1}^k \frac{1}{u_nu_{n+1}}.
\]
We can calculate as 
\begin{align*}
&1 -\sum_{n=0}^k \frac{1}{u_nu_{n+1}} -\frac{1}{u_1} +(u_1-1) \sum_{n=1}^k \frac{1}{u_nu_{n+1}} \\
&=1 - \frac{1}{u_1} - \frac{1}{u_0u_1} +(u_1-2) \sum_{n=1}^k \frac{1}{u_nu_{n+1}} \\
&=(u_1-2) \cdot \left( \frac{1}{u_1} +  \sum_{n=1}^k \frac{1}{u_nu_{n+1}} \right) >0
\end{align*}
When $i \ge 1$, we need to show 
\[
\frac{1}{u_i} +(u_{i-1} -u_i) \sum_{n=i}^k \frac{1}{u_nu_{n+1}}
>\frac{1}{u_{i+1}} +(u_{i}-u_{i+1}) \sum_{n=i+1}^k \frac{1}{u_nu_{n+1}}.
\]
Here, we have
\begin{align*}
&\frac{1}{u_i} +(u_{i-1}-u_i) \sum_{n=i}^k \frac{1}{u_nu_{n+1}} 
-\frac{1}{u_{i+1}} -(u_{i} -u_{i+1}) \sum_{n=i+1}^k \frac{1}{u_nu_{n+1}} \\
&=\frac{1}{u_i} -\frac{1}{u_{i+1}} + (u_{i-1}-u_i) \cdot \frac{1}{u_iu_{i+1}} 
+(u_{i+1} -2u_i +u_{i-1} ) \sum_{n=i+1}^k \frac{1}{u_nu_{n+1}} \\
&= (u_{i+1} -2u_i +u_{i-1} )\cdot \left( \frac{1}{u_iu_{i+1}} + \sum_{n=i+1}^k \frac{1}{u_nu_{n+1}}  \right)
>0
\end{align*}
which shows the assertion. \qed

When $a$ is an arc on the top side of the cycle $\gamma_i$ for $0\le i \le k$ and $\gamma_i(a)>0$,
\begin{align*}
\|(I-\Pi_k)\xi\| \cdot \varphi(a) &= 
\xi(a) - \frac{\sqrt{2(l+m)}}{\sqrt{m+3}} \sum_{n=0}^k \frac{1}{u_n u_{n+1}}
\sum_{s=0}^n (-1)^s u_s \gamma_s(a) \\
&=\frac{(-1)^{i+1}u_i}{\sqrt{2(m+3)}} 
\sum_{n=i}^k \frac{1}{u_n u_{n+1}}.
\end{align*}
When $\gamma_i(a)<0$, just multiply the above equation by $-1$.
The next proposition shows that the probability $q_i$ of finding the quantum walker at the top side
of $\gamma_i$ is also monotonically decrease.

\begin{theorem}\label{prop3.9.1}
Let $a_i$ be an arc on the top side of cycle $\gamma_i$ for $0\le i \le k$.
Then, the inequality 
\[
|\varphi(a_i)| > |\varphi(a_{i+1})|
\]
holds for $0 \le i \le k-1$.
\end{theorem}

\proof
It is enough to show
\[
\sum_{n=i}^k \frac{u_i}{u_n u_{n+1}} > \sum_{n=i+1}^k \frac{u_{i+1}}{u_nu_{n+1}}.
\]
By Lemma \ref{lem4.6} (2) and (3),
\begin{align*}
\sum_{n=i}^k \frac{u_i}{u_n u_{n+1}} - \sum_{n=i+1}^k \frac{u_{i+1}}{u_nu_{n+1}}
&=
\frac{u_i}{u_iu_{i+1}} - (u_{i+1} -u_i) \sum_{n=i+1}^k \frac{u_{i+1}}{u_nu_{n+1}} \\
&> \frac{1}{u_{i+1}}  - (u_{i+1} -u_i)\cdot \frac{1}{u_{i+1}} \cdot \frac{1}{u_{i+2}-u_{i+1}} \\
&=\frac{1}{u_{i+1}} \left( 1-  \frac{u_{i+1}-u_i}{u_{i+2} -u_{i+1}} \right) >0
\end{align*}
which shows the assertion. \qed

Proposition \ref{prop3.6.1} says that
the amplitude of the navigation vector may not be zero even at a point outside of the shortest route from the start to the goal.
On the other hand, Theorems \ref{prop3.8.1} and \ref{prop3.9.1} show that the quantum walker is more likely to be at a point on a shorter route.
Hence, in the case of ladder graph, the position of the quantum walker tells us a shorter route,
although not as much as in the case of tree graph.

\subsection{Comparison with the digraph with a sink at the goal}

A method of maze-solving was considered by adding the sink at the goal in \cite{MYLS}.
In this subsection, we discuss differences between the method in \cite{MYLS} and ours.

\begin{example}
The digraph in the left hand side of Figure {\rm \ref{fig3}} is considered in \cite{MYLS}.
When we put the sink at the goal, the state $(PU)^n \zeta$ does not converge.
The reason why this happens is that there exists an eigenvector in {\rm (1)} 
whose eigenvalue is not $-1$.
Let $\lambda = e^{\frac{2\pi {\rm i}}{3}}$ and $t=-\frac{\sqrt{3}}{2} {\rm i}$. 
Then, the vector defined as in the right hand side of Figure {\rm \ref{fig3}} is the eigenvector associated with the eigenvalue $\lambda$.
This eigenvector is the cause of oscillations of amplitudes (See \cite{KSS, MYLS}).

\begin{figure}[h]
\begin{center}
\includegraphics[bb=0 0 820 400, scale=0.4]{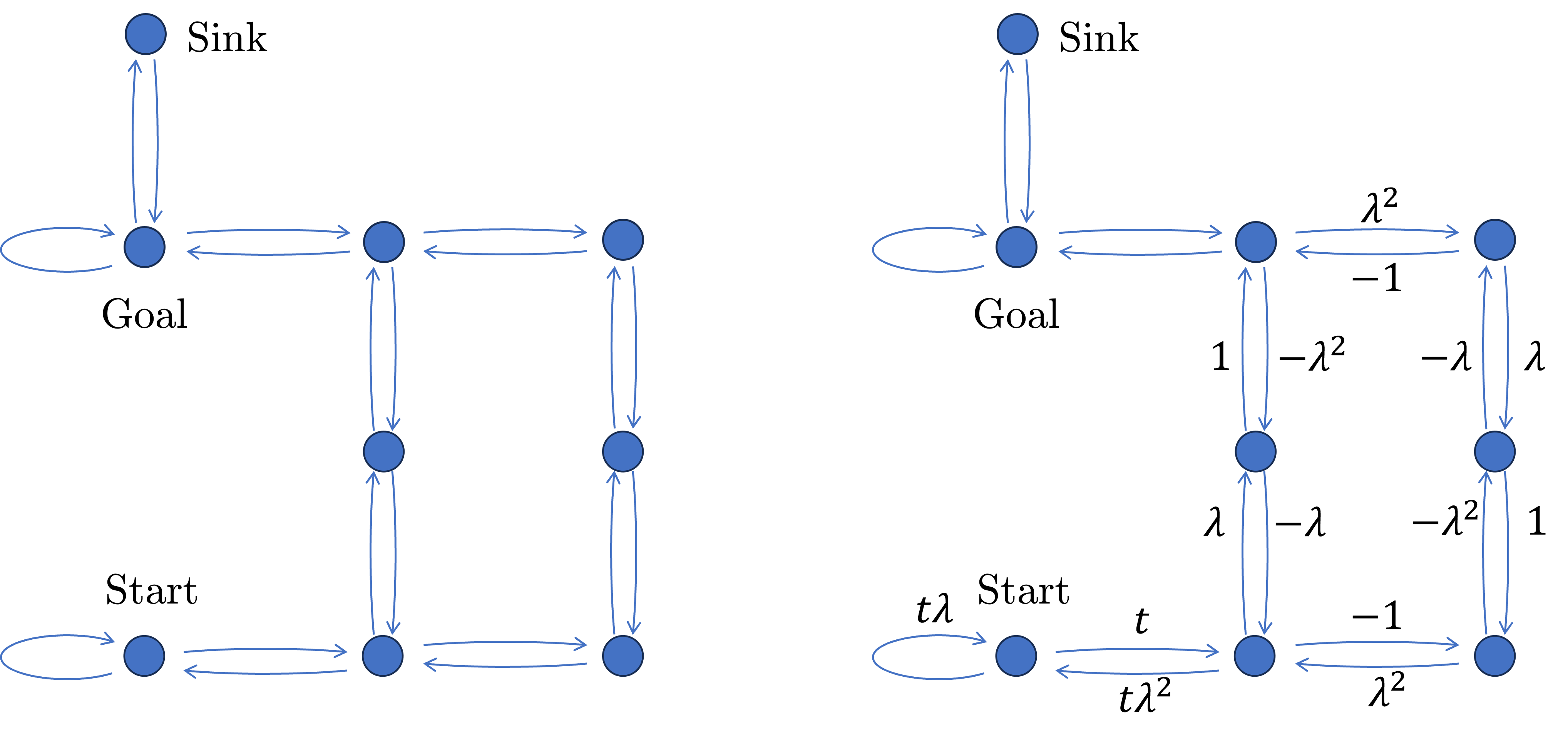}
\vspace*{-0.3cm}
\caption{Digraph considered in \cite{MYLS} and an eigenvector in (1).}\label{fig3}
\end{center}
\end{figure}

On the other hand, when we put the sink at the start,
the state $(-1)^n (PU)^n \zeta$ definitely converges. (See also Section \ref{sect4}.)
\end{example}

\begin{example}
When the sink is at the goal, we can make an example 
where the amplitudes of the state $(PU)^n \zeta$ 
appear at points other than the start-to-goal path.
Let  $\lambda =e^{\frac{\pi {\rm i}}{3}}$ and $k=6l+3$ $(l \in {\mathbb N})$.
Then, the vector defined as in Figure \ref{fig4} is in {\rm (1)} whose eigenvalue is not $-1$.
\begin{figure}[h]
\begin{center}
\includegraphics[bb=0 0 620 380, scale=0.4]{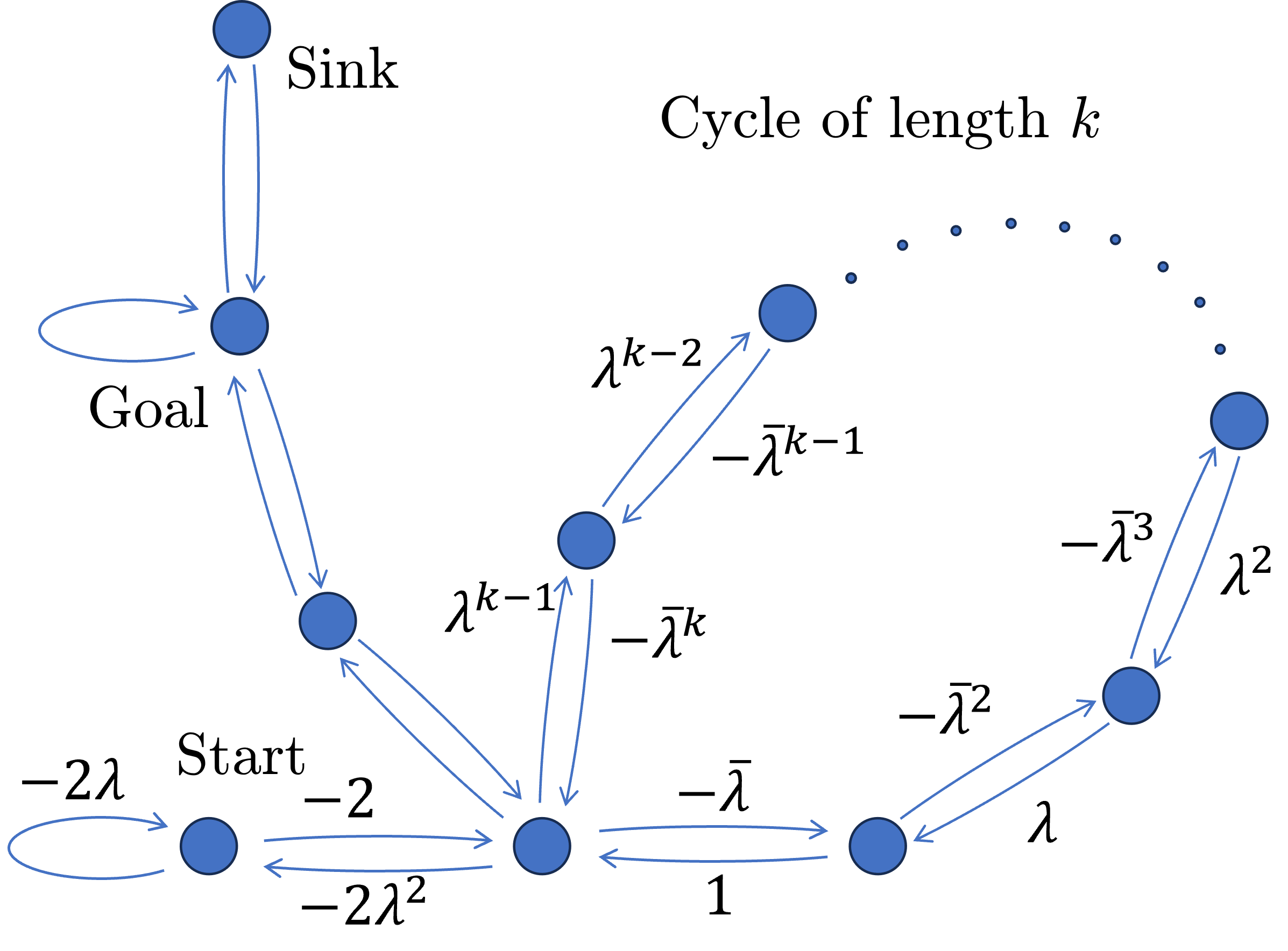}
\vspace*{-0.3cm}
\caption{An eigenvector in (1).}\label{fig4}
\end{center}
\end{figure}
Therefore, $(PU)^n \zeta$ always has non-zero coefficients on the cycle of length $k$.
This means that the position of the quantum walker does not tells us the shortest route.
Probably, it misleads us.
\end{example}

\section{Summary}
We have established a mathematical framework for maze solving using quantum walks and explicitly derived the limit distribution. Our method resembles the placement of food at the start and goal in slime mold maze-solving by incorporating self-loops at these vertices, which enables the emergence of the shortest path within the quantum walk structure. Moreover, we found that adding the sink at the start rather than at the goal effectively prevents the emergence of unstable states. These phenomena are based on the spectral properties of the Grover walk, characterized by a birth space defined by fundamental cycles, self-loops, and sinks. Although this eigenspace has not been utilized in existing quantum search algorithms, it plays an important role in our maze-solving methodology.

\if We proposed a potential of discrete-time quantum walks for a maze solving like the slime mold. We showed that Grover walk, which is a typical model for the discrete-time quantum walk, has the property choosing the shortest path between the start and goal autonomously in some special setting where there are many detours. 
We found that the feeds on the start and goal in the slime molds computing correspond to the self loops in the Grover walk maze solving. 
This phenomenon is based on the spectral property of the Grover walk having the {\it birth space} which is characterized by the fundamental cycles, self-loops and the sink. 
This eigenspace was wasted in the quantum search algorithms, however it plays the main role in the maze solving. \fi

\section*{Acknowledgments}

E.S. is supported by JSPS KAKENHI Grant Number 24K06863. H.O. is supported by JSPS KAKENHI Grant Number 23K03147.
\bibliographystyle{abbrv}
\bibliography{OMS_ref}

\end{document}